\documentclass[showpacs,twocolumn]{revtex4-1}
\usepackage[english]{babel}
\usepackage{graphicx}
\usepackage{epstopdf}
\usepackage{amsmath, amssymb}
\usepackage{dcolumn}
\usepackage{subfigure}

\DeclareMathOperator{\Tr}{Tr}

\begin{document}

\title{\bf Parameter estimation for a Lorentz invariance violation}

\author{H. A. S. Costa}\email[]{hascosta@ufpi.edu.br}
\author{P. R. S. Carvalho}\email[prscarvalho@ufpi.edu.br]{ }
\author{I. G. da Paz}\email[irismarpaz@ufpi.edu.br]{ }

\affiliation{Departamento de F\'{\i}sica, Universidade Federal do Piau\'{\i}, Campus Ministro Petr\^{o}nio Portela, 64049-550, Teresina, PI, Brazil}


\begin{abstract}

We employ techniques from quantum estimation theory (QET) to estimate the Lorentz violation parameters in the 1+3-dimensional flat spacetime. We obtain and discuss the expression of the quantum Fisher information (QFI) in terms of the Lorentz violation parameter $\sigma_0$ and the momentum k of the created particles. We show that the maximum QFI is achieved for a specific momentum $k_{\mathrm{max}}$. We also find that the optimal precision of estimation of the Lorentz violation parameter is obtained near the Planck scale.

\end{abstract}

\maketitle

\section{Introduction}

  Invariance under Lorentz transformation is a fundamental symmetry in the General Relativity and in the Standard Model of particle physics. However it is well known that at energies near the Planck scale, $M_p \approx 1.22\times 10^{19}$ GeV, the effects of quantum gravity are important and the Lorentz symmetry is violated. The Lorentz symmetry breaking mechanism has been initially discussed in the high energy context of string theory \cite{Green, Kostelecky, Polshinski}. More recently the possible effects of Lorentz violation has been studied in the contexts of loop quantum gravity \cite{Rovelli}, inflationary cosmology \cite{Brandenberg1, Brandenberg2}, dark energy problem \cite{Bertolami1, Bertolami2}, baryogenesis \cite{Carroll, Alfaro}, gravitational waves \cite{Mirshekari} etc. In addition, several studies of the phenomenological aspects of quantum gravity suggest a violation of Lorentz symmetry due to some deviations from the standard dispersion relation \cite{Greisen, Antonov, Aloisio, Jacobson, Mattingly, Camelia}.

  The leading order modification to the dispersion relation can be represented by
  \begin{align} \label{E}
  E^2 = p^2c^2 + m^2c^4 + \Lambda p^nc^n,
  \end{align}
 where $E$, $p$ and $m$ are the energy, momentum and mass of the particle, respectively. The parametrs $\Lambda$ and $n$  characterize the Lorentz violation mechanism ($n$ is dimensionless and $\Lambda$ has dimension of $[\mathrm{mass}]^{2-n}$). Different approaches to quantum gravity suggest different dispersion relations from Eq. (\ref{E}). As examples of such models we can mention the Double Special Relativity Theory \cite{GAC1, Magueijo, GAC2, GAC3}, the Extra-Dimensional Theories \cite{Sefiedgar}, the Horava-Lifshitz Theory \cite{Horava, Horava, Vacaru, Garattini1} and the Theories with Non-Commutative Geometries \cite{Garattini1, Garattini2, Garattini3}. Another path pursued in recent research is the possibility of particle creation as well as entropy generation due to Lorentz invariance violation in the course of cosmological evolution \cite{Khajeh, Khosravi, Mohammadzadeh}. In this viewpoint, the existence of particles and the entropy production can be interpreted as an evidence for quantum gravity effects in the early universe. Since the Lorentz symmetry violation mechanism has an important role in modern theoretical physics, these investigations are of great theoretical and experimental interest.

 To obtain a better understanding of the mechanisms of Lorentz violation in the early universe, it is interesting to study the dynamics of Lorentz violation from the point of view of quantum estimation theory (QET) \cite{Helstrom, Holevo, Giovanneti, Paris}. Motivated by this consideration, in this work we consider a time dependent deformed dispersion relation to describe the effects of Lorentz invariance violation in the particle creation process. Our interest is to investigate the higher bound put on how precisely we are allowed to estimate the Lorentz violation parameter. In this intuit, we apply techniques from QET to determine the lower bound in an estimation scheme. This lower bound, known as Cramer-Rao inequality is given by the inverse of the so-called Fisher information $F(\theta)$:
  \begin{align}
   (\bigtriangleup\theta)^2 \geq \frac{1}{F(\theta)},
\end{align}
 where
 \begin{align} \label{Fc}
 F(\theta) = \int d\xi P(\xi;\theta)\left[\frac{\partial}{\partial\theta}\log P(\xi;\theta)\right]^2.
 \end{align}
Here $\theta$ is the parameter to be estimated, $\xi$ denotes the outcome of a measurement and $P(\xi;\theta)$ is the probability distribution to obtain a certain measurement result $\xi$ parametrized by $\theta$. In particular, we analytically calculate
the quantum Fisher information (QFI) and  derive the lower bound for the precision of Lorentz violation parameter. We find that the QFI approaches its maximum value for a specific momentum, $k_{\mathrm{max}}$, which allows to archive the lower bound for precision of the estimation of Lorentz violation parameter. In addition, we found that near the Planck scale the Cramer-Rao inequality saturates producing the small possible  value for the variance $\bigtriangleup\sigma_0$, which is the ultimate lower bound of precision.

This paper is organized as follow: in Sec. II, we introduce the dynamical Lorentz violation model in flat spacetime and obtain the reduced density matrix. In Sec. III we evaluate the QFI and the corresponding bounds for precision of Lorentz violation parameter. Finally, a brief summary of the main results obtained is presented in Sec. IV.

\section{The model}

  It is generally believed that the quantum gravity effects are dominant in the very early universe. An approach that has been extensively considered in the literature to study such effects are to employ models based on modiﬁed dispersion relations. These models effectively introduce a fundamental scale (Planck scale) beyond which the modiﬁcations to the linear dispersion relation become important. In particular, the model for a modification in the dispersion relation which we use to derive our results reads
\begin{align} \label{drLV}
\omega_k^2 = k^2 - \sigma^2k^4,
\end{align}
 where the coefficient $\sigma^2$ represents the quantum gravity parameter and expected to be of the order of Planck scale. Notice that the last term of the modified dispersion relation (\ref{drLV}) breaks the Lorentz invariance. Furthermore, it is worth mentioning that this modification in the dispersion relation is not a unique choice. In particular, introducing a cubic term has been discussed in Refs. 15 and 36 and a more general dispersion relation in Ref. 17. Since the Lorentz violation effects are expected to appear at early time and then entirely subside at late time, we will assume that the parameter $\sigma$ has a time dependence such that for very early times is a constant value and vanishing for late times. In natural units ($\hbar = c = 1$), when the values of $\sigma$ approach 1 from of left, one recovers prediction of certain quantum gravity models.

 The action that can lead the modified dispersion relation (\ref{drLV}) for a scalar field $\phi$ in a 3+1-dimensional Minkowski spacetime is given by \cite{Jacobson2}
\begin{align}
  S = \frac{1}{2}\int d^4x[\eta^{\mu\nu}\partial_{\mu}\phi\partial_{\mu}\phi + \sigma^2(D^2\phi)^2], \nonumber
\end{align}
  where $\eta_{\mu\nu}$ is the metric tensor and $\sigma$ is the time dependent Lorentz violation parameter. The term $D^2\phi$ corresponds to the covariant form of the spatial Laplacian defined as
  \begin{align}
  D^2\phi = -q^{\mu\nu}\partial_{\nu}(q_{\mu}^{\tau}\partial_{\tau}\phi),
\end{align}
  where $q_{\mu\nu}$
  \begin{align*}
  q_{\mu\nu} = -\eta_{\mu\nu} + u_{\mu}u_{\nu}, \quad \eta^{\mu\nu}u_{\mu}u_{\nu} = 1,
\end{align*}
   indicates the spatial metric orthogonal to the unit timelike vector $u^{\mu}$. The vector field $u^{\mu}$ can be interpreted as the four velocity of a preferred inertial observer. In the rest frame of this preferred observer we have $u^{\mu} = (1, 0, 0, 0)$.
   
  In the preferred frame the dynamics of the field $\phi$ is governed by the Klein-Gordon equation
  \begin{align} \label{KG}
  (\eta^{\mu\nu}\partial_{\mu}\partial_{\nu} - \sigma^2q^{\mu\nu}q^{\alpha\beta}\partial_{\mu}\partial_{\nu}\partial_{\alpha}\partial_{\beta})\phi(\textbf{x}, t) = 0.
\end{align}
   Since Minkowski spacetime admits a global timelike Killing vector $\partial_t$, the corresponding energy conservation provides a natural way of classifying the solutions of Eq.(\ref{KG}) into positive and negative frequencies. The solutions of Eq. (\ref{KG}) are the eigenfunctions $u_{\textbf{k}}$ and $u_{\textbf{k}}^{*}$ of the operator $i\mathcal{L}_{\partial_t}$ with positive and negative eigenvalues, respectively:
  \begin{align}
  i\mathcal{L}_{\partial_t}u_{\textbf{k}} &= \omega_ku_{\textbf{k}}, \quad (\text{postive frequency}), \\
  i\mathcal{L}_{\partial_t}u_{\textbf{k}}^* &= -\omega_ku_{\textbf{k}}^*, \quad (\text{negative frequency}).
\end{align}
 Here $\mathcal{L}$ denotes Lie derivative defined as $\mathcal{L}_{\xi}u_{\textbf{k}} = \xi^{\mu}\partial_{\mu}u_{\textbf{k}}$. Thus, we can expand an arbitrary solution of Eq. (\ref{KG}) as a sum of these positive and negative frequencies solutions:
  \begin{align}
  \hat{\phi}(\textbf{x}, t) = \int d^3\textbf{k}(\hat{a}_{\textbf{k}}u_{\textbf{k}} + \hat{a}_{\textbf{k}}^{\dagger}u_{\textbf{k}}^{*}),
\end{align}
  where $\hat{a}_{\textbf{k}}^{\dagger}$ and $\hat{a}_{\textbf{k}}$ are creation and annihilation operators, respectively, which satisfy the following commutation relations
  $$[\hat{a}_{\textbf{k}}, \hat{a}_{\textbf{k}'}] = [\hat{a}_{\textbf{k}}^{\dagger}, \hat{a}_{\textbf{k}'}^{\dagger}] = 0, \quad [\hat{a}_{\textbf{k}}, \hat{a}_{\textbf{k}'}^{\dagger}] = \delta_{\textbf{k}\textbf{k}'}.$$
 
\begin{figure}[htp]
    \centering
    \includegraphics[scale=0.5]{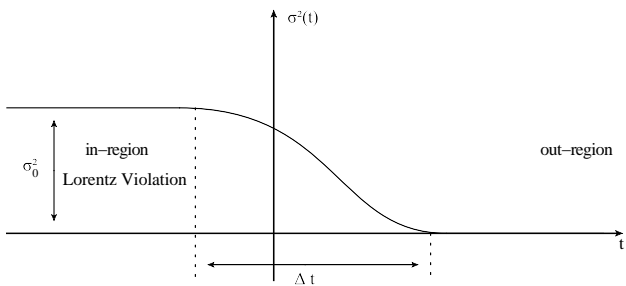}
    \caption{Illustration of the time evolution of the Lorentz violation parameter $\sigma^2(t)$ which possesses two asymptotic regions. The parameter $\sigma_0$ is the initial value of the Lorentz violation parameter and represents the quantum gravity regime in the early universe.}
    \label{fig0}
\end{figure}

 The vacuum state of the Lorentz violation model in Minkowski spacetime can be defined as
\begin{align}
\hat{a}_{\textbf{k}}|0^{\mathrm{LV}}\rangle = 0, \quad \forall \textbf{k}.
\end{align}
 Note that the vacuum state $|0^{\mathrm{LV}}\rangle$ was defined to a coordinate system which corresponds to the rest frame of a preferred inertial observer. Due to the linearity in $\phi$, the solutions $u_{\textbf{k}}$ and their complex conjugates form an orthonormal basis of solutions with respect to the Klein-Gordon scalar product
  \begin{align} \label{KGSP}
  (u_{\textbf{j}}, u_{\textbf{k}}) = -i\int d\Sigma n^{\mu}(u_{\textbf{j}}\overleftrightarrow{\partial}_{\mu}u_{\textbf{k}}^* - \sigma^2q^{\mu\nu}q^{\alpha\beta}u_{\textbf{j}}\overleftrightarrow{\partial_{\mu}\partial_{\nu}\partial_{\alpha}}u_{\textbf{k}}^*), \nonumber
  \end{align}
  where $d\Sigma$ is the volume element and $n^{\mu}$ is a future-directed unit vector orthogonal to $\Sigma$. The notations $\overleftrightarrow{\partial}_{\mu}$ and $\overleftrightarrow{\partial_{\mu}\partial_{\nu}\partial_{\alpha}}$ in the Eq. (\ref{KGSP}) are defined by
  \begin{align*}
  u_{\textbf{j}}\overleftrightarrow{\partial}_{\mu}u_{\textbf{k}} &= u_{\textbf{j}}\partial_{\mu}u_{\textbf{k}} - u_{\textbf{k}}\partial_{\mu}u_{\textbf{j}}, \\
  u_{\textbf{j}}\overleftrightarrow{\partial_{\mu}\partial_{\nu}\partial_{\alpha}}u_{\textbf{k}} &= u_{\textbf{j}}\partial_{\mu}\partial_{\nu}\partial_{\alpha}u_{\textbf{k}} - u_{\textbf{k}}\partial_{\mu}\partial_{\nu}\partial_{\alpha}u_{\textbf{j}} \\
  & - \partial_{\mu}u_{\textbf{j}}\partial_{\nu}\partial_{\alpha}u_{\textbf{k}} + \partial_{\mu}u_{\textbf{k}}\partial_{\nu}\partial_{\alpha}u_{\textbf{j}} \\
  & + \partial_{\nu}u_{\textbf{j}}\partial_{\mu}\partial_{\alpha}u_{\textbf{k}} - \partial_{\nu}u_{\textbf{k}}\partial_{\mu}\partial_{\alpha}u_{\textbf{k}} \\
  & - \partial_{\alpha}u_{\textbf{j}}\partial_{\nu}\partial_{\mu}u_{\textbf{k}} + \partial_{\alpha}u_{\textbf{k}}\partial_{\nu}\partial_{\mu}u_{\textbf{j}},
  \end{align*}
  respectively. Since the spacetime is homogeneouns, the mode functions $u_{\textbf{k}}$ can be separated as
  \begin{align}
  u_{\textbf{k}}(\textbf{x}, t) = (2\pi)^{-\frac{3}{2}}e^{i\textbf{k}\cdot\textbf{x}}\chi_{\textbf{k}}(t),
\end{align}
  where $\chi_{\textbf{k}}(t)$ satisfy the following equations of motion
  \begin{align}\label{EqM}
   \frac{d^2}{dt^2}\chi_{\textbf{k}}(t) + (k^2 - \sigma^2k^4)\chi_{\textbf{k}}(t) = 0.
  \end{align}
  Let us assume that the time evolution of the Lorentz violation parameter $\sigma^2$ is given by
  \begin{align} \label{sigma}
  \sigma^2(t) = \frac{\sigma_0^2}{2}[1 - \tanh(\rho_0 t)],
\end{align}
  where $\sigma_0$ is the initial value of the Lorentz violation parameter and represents the quantum gravity regime in our model. $\rho_0$ is a positive real parameter controlling the reduction of the effects of Lorentz violation in the time interval $\Delta t$ shown in Fig. 1. From Eq. (\ref{sigma}) we can find that the parameter $\rho_0$ is proportional to the inverse of the time interval $\Delta t$. Thus, $\rho_0$ can be interpreted as the rate of reduction of the trans-Planckian effects during the early-stage evolution of the universe. We assume a flat universe throughout the present paper, so the Minkowski coordinates $(t, \textbf{x})$ are the comoving coordinates\footnote{Comoving coordinates in a reference frame are coordinates so that the reference observer of the frame have constant spatial coordinates} of an inertial reference frame (the cosmic rest frame) in an expanding universe model. The time $t$ is the proper time of an observer at rest and it is related to the conformal time $\eta$ via $t = a\eta$, where $a$ is the scale factor describing cosmic expansion. We would like to emphasize that this form of the time evolution of the Lorentz violation parameter is chosen such that the resulting equations are exactly solvable. In addition, our model is equivalent to the model suggested in Refs. 30 and 31. The Lorentz violation parameter $\sigma^2(t)$ is sufficiently smooth and approaches constant values in the distant past $\sigma^2(t \rightarrow -\infty) = \sigma_0^2$ and in the far future $\sigma^2(t \rightarrow \infty) = 0$ as depicted in figure (\ref{fig0}). In such asymptotic regions, the dispersion relation in Eq. (\ref{drLV}) will take different forms. For the very early times it has the form $\omega_k^{\mathrm{in}} = \sqrt{k^2 - \sigma_0^2k^4}$ and for the very late times the dispersion relation becomes the standard one, namely $\omega_k^{\mathrm{out}} = \sqrt{k^2}$.

  Using the Eq. (\ref{sigma}), we can solve the Eq. (\ref{EqM}) in terms of hypergeometric functions and obtain the following results for the normalized modes in the in-region and in the out-region. They are
   \begin{align}
  u_{\textbf{k}}^{\mathrm{in}}(\textbf{x}, t) &= \chi^{\mathrm{in}}(t)e^{i\textbf{k}\cdot\textbf{x} - i\omega_{+}t -\frac{i\omega_{-}}{\rho_0}\ln[2\cosh(\rho_0 t)]}, \\
  u_{\textbf{k}}^{\mathrm{out}}(\textbf{x}, t) &= \chi^{\mathrm{out}}(t)e^{i\textbf{k}\cdot\textbf{x} - i\omega_{+}t -\frac{i\omega_{-}}{\rho_0}\ln[2\cosh(\rho_0 t)]},
  \end{align}
  respectively, where
  \begin{align*}
  \chi^{\mathrm{in}}(t) &= \frac{1}{\sqrt{4\pi\omega_{k}^{\mathrm{in}}}}F[a, b; c_1; \frac{1}{2}(1 + \tanh(\rho_0 t))], \\
  \chi^{\mathrm{out}}(t) &= \frac{1}{\sqrt{4\pi\omega_{k}^{\mathrm{out}}}}F[a, b; c_2; \frac{1}{2}(1 - \tanh(\rho_0 t))].
  \end{align*}
  The constants above are defined as
  \begin{align*}
  a &= 1 + \frac{i\omega_{-}}{\rho_0}, \quad
  b = \frac{i\omega_{-}}{\rho_0}, \\
  c_1 &= 1 - \frac{i\omega_k^{\mathrm{in}}}{\rho_0}, \quad
  c_2 = 1 + \frac{i\omega_k^{\mathrm{out}}}{\rho_0},
  \end{align*}
  and the frequencies as
  \begin{align*}
  \omega_{\pm} = \frac{1}{2}(\omega_k^{\mathrm{out}} \pm \omega_k^{\mathrm{in}}).
  \end{align*}
  The asymptotic solutions are connected by a Bogoliubov transformation that only mixes the modes of the same momentum $k$:
  \begin{align}
  u_{\textbf{k}}^{\mathrm{in}}(\textbf{x}, t) = \alpha_{k}u_{\textbf{k}}^{\mathrm{out}}(\textbf{x}, t) + \beta_{k}u_{-\textbf{k}}^{\mathrm{out}*}(\textbf{x}, t),
  \end{align}
  which accordingly implies a transformation between the creation and annihilation operators as
  \begin{align} \label{a}
  \hat{a}_{\textbf{k}}^{\mathrm{in}} = \alpha_k\hat{a}_{\textbf{k}}^{\mathrm{out}} + \beta_k\hat{a}_{-\textbf{k}}^{\mathrm{out}\dagger}
  \end{align}
  where the Bogoliubov coefficients $\alpha_{k}$ and $\beta_{k}$ satisfy the normalization condition $|\alpha_{k}|^2 - |\beta_{k}|^2 = 1$. In this particular case they are readily evaluated to
  \begin{align}\label{BC}
     \alpha_{k} &= \sqrt{\frac{\omega_{k}^{\mathrm{out}}}{\omega_{k}^{\mathrm{in}}}}\frac{\Gamma(1 - \frac{i\omega_{k}^{\mathrm{out}}}{\rho_0})\Gamma(\frac{-i\omega_{k}^{\mathrm{in}}}{\rho_0})}{\Gamma(\frac{-i\omega_{+}}{\rho_0})\Gamma(1 - \frac{i\omega_{+}}{\rho_0})}, \\
     \beta_{k} &= \sqrt{\frac{\omega_{k}^{\mathrm{out}}}{\omega_{k}^{\mathrm{in}}}}\frac{\Gamma(1 - \frac{i\omega_{k}^{\mathrm{out}}}{\rho_0})\Gamma(\frac{i\omega_{k}^{\mathrm{in}}}{\rho_0})}{\Gamma(\frac{i\omega_{-}}{\rho_0})\Gamma(1 + \frac{i\omega_{-}}{\rho_0})}.
\end{align}
   Now let us consider the vacuum state in the asymptotic past $|0^{\mathrm{LV}}\rangle_{\mathrm{in}}$, which is annihilated by $\hat{a}_{\textbf{k}}^{\mathrm{in}}$. We would like to know the form of the state $|0^{\mathrm{LV}}\rangle_{\mathrm{in}}$ in the asymptotic future. To this end, we use the fact that $\hat{a}_{\textbf{k}}^{\mathrm{in}}|0^{\mathrm{LV}}\rangle_{\mathrm{in}} = 0$. If we substitute $\hat{a}_{\textbf{k}}^{\mathrm{in}}$ in terms of `out' operators using Eq. (\ref{a}), we obtain
  \begin{align} \label{Eqinout}
  (\alpha_k\hat{a}_{\textbf{k}}^{\mathrm{out}} + \beta_k\hat{a}_{-\textbf{k}}^{\mathrm{out}\dagger})|0^{\mathrm{LV}}\rangle_{\mathrm{in}} = 0.
  \end{align}
  Without loss of generality, we consider the following ``ansatz" for the vacuum state in the asymptotic future
  \begin{align}
  |0^{\mathrm{LV}}\rangle_{\mathrm{in}} = \sum_{n = 0}^{\infty}A_n|n_k, n_{-k}\rangle_{\mathrm{out}}.
\end{align}
   By substituting the expression above into Eq. (\ref{Eqinout}), we get the recurrence relation $A_n = \left(\frac{\beta_k^*}{\alpha_k^*}\right)A_0$ and from the normalization condition we find $|A_0|^2 = 1 - \left|\frac{\beta_k}{\alpha_k}\right|^2$. Therefore, the vacuum state $|0^{\mathrm{LV}}\rangle_{\mathrm{in}}$ in terms of `out' modes in the asymptotic future becomes
    \begin{align}
    |0^{\mathrm{LV}}\rangle_{\mathrm{in}} = \sqrt{1 - \gamma_{k}}\sum_{n = 0}^{\infty}\gamma_{k}^n|n_{k}, n_{-k}\rangle , \label{0k}
    \end{align}
    where
    \begin{align} \label{Gammak}
    \gamma_{k} = \left|\frac{\beta_k}{\alpha_k}\right|^2 = \frac{\sinh^2(\frac{\pi}{\rho_0}\omega_{-})}{\sinh^2(\frac{\pi}{\rho_0}\omega_{+})}.
    \end{align}
    Since we are working with a single mode, we will drop the frequency index $k$. Note that the state in Eq. (\ref{0k}) is a \textit{bona fide} two mode squeezed state. After particle creation in the out-region, each pair of modes with opposite momenta are separated on cosmological scale, thus a local inertial observer in the out-region access only one of each pair of modes, say $k$, with the mode $-k$ traced out. This means that an inertial observer in the out-region detects a distribution of particles with modes $k$ according to the marginal state
    \begin{align}
 \hat{\rho}_{k} &= \Tr_{-k}[\hat{\rho}_{k, -k}], \nonumber \\
                &= (1 - \gamma_{k})\sum_{n = 0}^{\infty}\gamma_{k}^n|n_k\rangle\langle n_k|,
\end{align}
   where the density matrix of whole state $\hat{\rho}_{k,-k}$ corresponds to $\hat{\rho}_{k,-k} = |0^{\mathrm{LV}}\rangle_{\mathrm{in}}\langle 0^{\mathrm{LV}}|$. Note that information about the parameters $\sigma_0$ and $\rho_0$ is codified in the final state $\hat{\rho}_{k}$. By applying quantum estimation techniques on the state $\hat{\rho}_{k}$ we can find a set of measurements that allows us to estimate the parameters $\sigma_0$ and $\rho_0$ with higher precision. This means that we can achieve the ultimate bound of parameters $\sigma_0$ and $\rho_0$ using measurement schemes feasible with current technology, e.g., the precision of the estimation can be improved by choosing the projective measurements corresponding to eigenvectors of the particle states. In the next section, our main aim is to estimate the Lorentz violation parameter $\sigma_0$ with the minimum variance  $\triangle\sigma_0$, which is the ultimate bound of precision imposed by the quantum theory.

  \section{Optimal estimation of the Lorentz violation parameter}

  In this section, our aim is to investigate how precisely one can estimate the Lorentz violation parameters. Specifically, we will estimate the parameter $\sigma_0$ based on the Cramér-Rao inequality. Thus we should at first evaluate the QFI in terms of the Bogoliubov coefficients $\alpha_k$ and $\beta_k$ which encodes information on the estimated parameter.

  The quantum analogue of the Fisher information (\ref{Fc}) is defined as
  \begin{align}
  F(\theta) = \int d\xi \frac{\mathrm{Re}(\Tr[\hat{\rho}\mathrm{E}_{\xi}\hat{L}_{\theta}])}{\Tr[\hat{\rho}\mathrm{E}_{\xi}]},
\end{align}
  where $\mathrm{E}_{\xi}$ are the elements of a positive operator-valued measure(POVM), $\hat{\rho}$ is the density matrix parametrized by the quantity $\theta$, and $\hat{L}_{\theta}$ is a self-adjoint operator defined by
  \begin{align}
   \partial_{\theta}\rho = \frac{1}{2}[\hat{L}_{\theta}\hat{\rho} + \hat{\rho}\hat{L}_{\theta}].
   \end{align}
  The Fisher information can be shown to be upper bounded by the QFI \cite{Helstrom2, Braunstein}
  \begin{align}
  F(\theta) \leq H(\theta) \equiv \Tr[\hat{\rho}\hat{L}_{\theta}].
  \end{align}
  In order to estimate the parameter $\sigma_0$ we need evaluated the QFI of the particle state $\hat{\rho}_k$ parametrized by the quantity $\sigma_0$, i.e., we need calculated
  \begin{align} \label{F0}
  H(\sigma_0) = \Tr[\hat{\rho}\hat{L}_{\sigma_0}].
  \end{align}
  Since the density matrix $\hat{\rho}_k$ has a spectral decomposition, Eq. (\ref{F0}) can be rewritten as
  \begin{align} \label{F}
  H(\sigma_0) &= \sum_{n = 0}^{\infty}\frac{1}{\lambda_n}\left(\partial_{\sigma_0}\lambda_n\right)^2 \nonumber \\
  &+ 2\sum_{n\neq m = 0}^{\infty}\frac{(\lambda_m - \lambda_n)^2}{\lambda_m + \lambda_n}|\langle n_k|\partial_{\sigma_0}m_k\rangle|^2,
  \end{align}
  where $\lambda_{n} = (1 - \gamma_{k})\gamma_{k}^n$. Note that the first term in Eq. (\ref{F}) represents the classical Fisher information and the last term contains the truly quantum contribution. This term does not contribute to $H(\sigma_0)$ because $|n_k\rangle$ and $|m_k\rangle$ with $n\neq m$ locate on different subspaces, i.e., $\langle n_k|\partial_{\sigma_0}m_k\rangle = 0$ $\forall n, m$. We can obtain the analytical expression for the classical part of $H(\sigma_0)$ by introducing the Eq. (\ref{Gammak}):
  \begin{align}
  H_c(\sigma_0) &= \frac{1}{1 - \gamma_k}\sum_{n=0}^{\infty}\frac{1}{\gamma_k^n}\left[\partial_{\sigma_0}(1 - \gamma_k)\gamma_k^n\right]^2 \nonumber \\
  &= \frac{(\partial_{\sigma_0}\gamma_k)^2}{1 - \gamma_k}\sum_{n=0}^{\infty}\gamma_k^n - 2\partial_{\sigma_0}\gamma_k\sum_{n=0}^{\infty}\partial_{\sigma_0}\gamma_k^n \nonumber \\
  & + (1 - \gamma_k)\sum_{n=0}^{\infty}\frac{(\partial_{\sigma_0}\gamma_k^n)^2}{\gamma_k^n}.
  \end{align}
Using the relations
   \begin{align}
   \partial_{\sigma_0}\gamma_k &= -\frac{\pi k^4\sigma_0}{\omega^{\mathrm{in}}_k\rho_0}\left[\frac{1}{\tanh(\frac{\pi\omega_{-}}{\rho_0})} + \frac{1}{\tanh(\frac{\pi\omega_{+}}{\rho_0})}\right]\gamma_k, \nonumber \\
   \partial_{\sigma_0}\gamma_k^n &= -\frac{\pi k^4\sigma_0}{\omega^{\mathrm{in}}_k\rho_0}\left[\frac{1}{\tanh(\frac{\pi\omega_{-}}{\rho_0})} + \frac{1}{\tanh(\frac{\pi\omega_{+}}{\rho_0})}\right]n\gamma_k^n, \nonumber
   \end{align}
  the classical part of $H(\sigma_0)$ is found to be
  \begin{align} \label{F2}
  H_c(\sigma_0) = \frac{\gamma_k^2}{(1 - \gamma_k)^2}\frac{\pi^2 k^8\sigma_0^2}{\omega^{\mathrm{in}2}_k\rho_0^2}\left[\frac{\tanh(\frac{\pi\omega_{-}}{\rho_0}) + \tanh(\frac{\pi\omega_{+}}{\rho_0})}{\tanh(\frac{\pi\omega_{-}}{\rho_0})\tanh(\frac{\pi\omega_{+}}{\rho_0})}\right]^2.
  \end{align}
  Notice that $H_c(\sigma_0)$ depends on the parameters $\sigma_0$ and $\rho_0$. Thus, to find out the optimal working regimes we have to maximize $H_c(\sigma_0)$ over these two parameters. Since we assume that the Lorentz violation mechanism occurs in the scale of Planck, the value of parameter $\sigma_0$ has to approach 1 from the left, i.e., $\sigma_0 \longmapsto 1^{-}$.
 
  \begin{figure}[htp]
    \centering
    \includegraphics[scale=0.21]{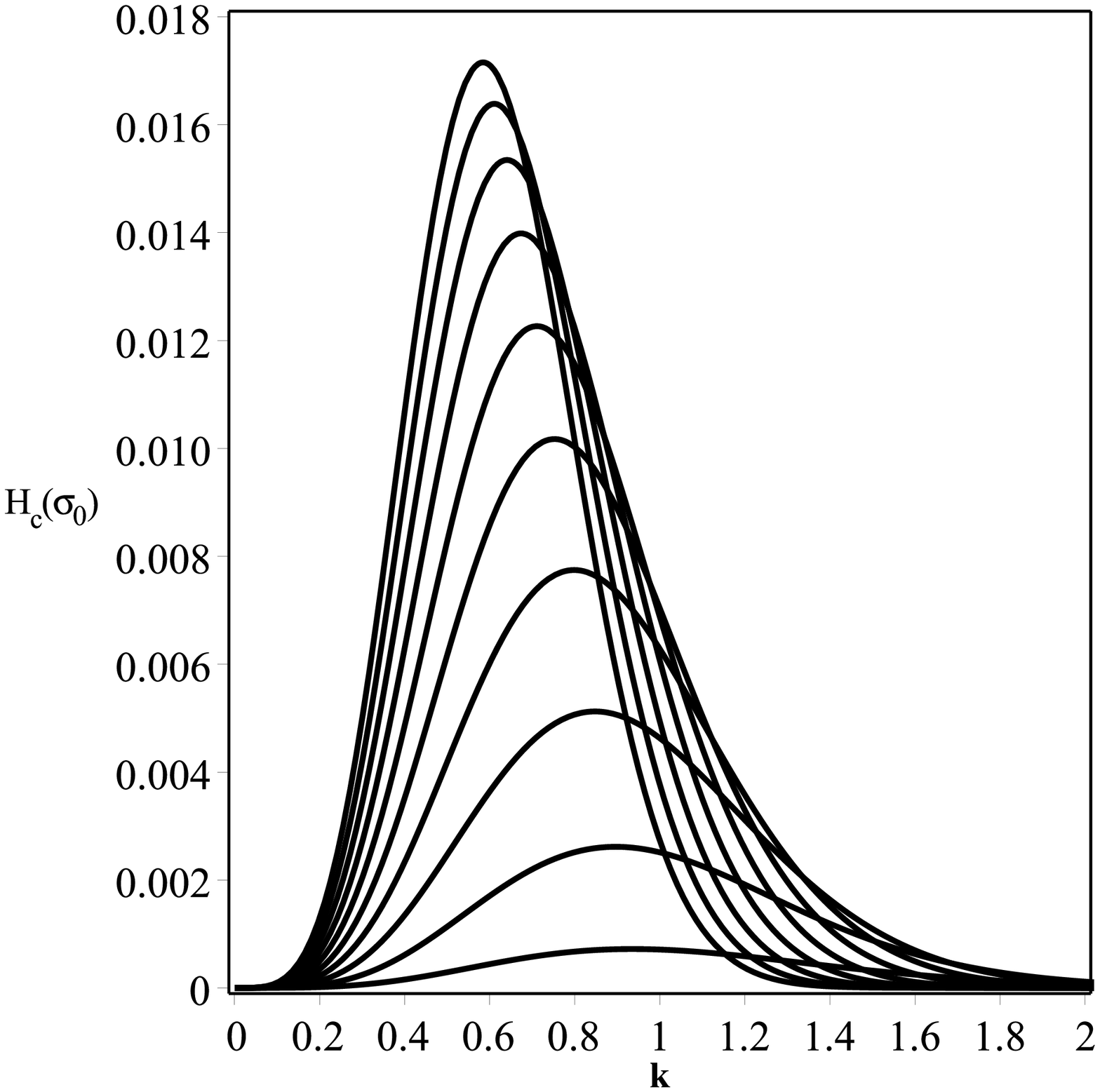} \includegraphics[scale=0.21]{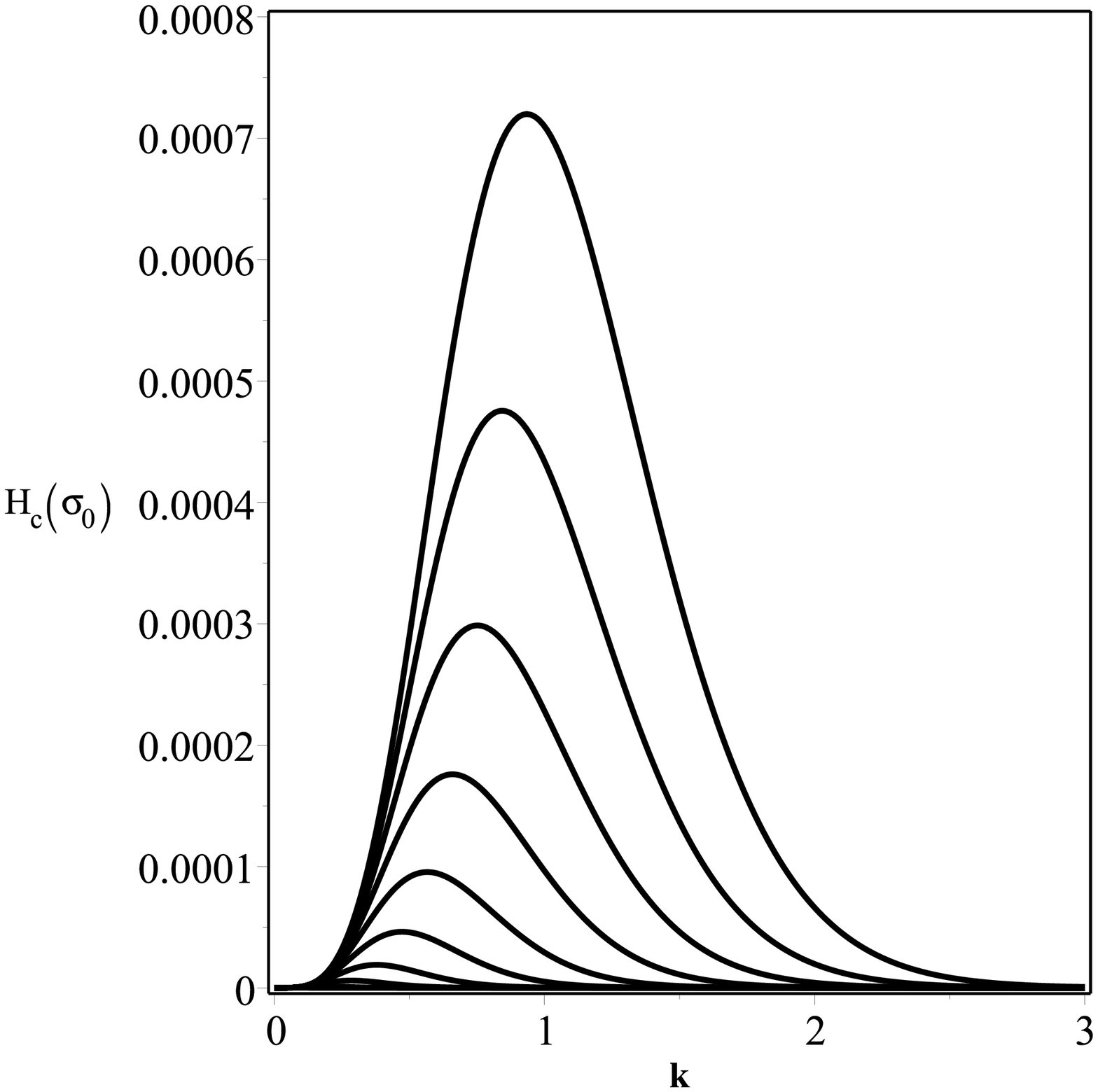}
    \caption{ Classical part of $H(\sigma_0)$ as function of the $k$ for different values of the Lorentz violation parameter $\sigma_0 = 0.1, ..., 1.0$ with $\rho_0 = 1$ (left) and for different of the rate of reduction $\rho_0 = 0.1, ..., 1.0$ with $\sigma_0 = 0.1$ (right). The higher spectral peaks in (left) and (right) correspond to $\sigma_0 = 1$ and $\rho_0 = 1$, respectively.}
    \label{fig2}
\end{figure}

  In Fig. (\ref{fig2}) (left) we plot $H_c(\sigma_0)$  for different values of $\sigma_0$ as a function of the momentum $k$ of the created particles. We can see that $H_c(\sigma_0)$  approaches its maximum value for a specific momentum, $k_{\mathrm{max}}$. This means that modes with this characteristic frequency are more sensitive to the Lorentz invariance violation. In other words, modes with this characteristic frequency are more easily excited by dynamics of Lorentz violation mechanism. In addition, this special behavior form the characteristic peak that $H_c(\sigma_0)$  presents at a specific momentum encodes information about the optimal value of $\sigma_0$. Fig. (\ref{fig2}) (left) shows also that the characteristic peak increases and shifts to the left as $\sigma_0$ grows. To illustrate this result, we show in Table I the relation between $k_{\mathrm{max}}$ and $\sigma_0$. In Fig. (\ref{fig2}) (right) we also can observe that the higher spectral peaks of $H_c(\sigma_0)$ increase and shift to the right when the rate of reduction $\rho_0$ grows. This results suggest that a regime of non-adiabatic  reduction ($\rho_0$ large) favouring the encoding of information about the effects of Lorentz violation in early times.

\begin{figure}[htp]
    \centering
    \includegraphics[scale=0.21]{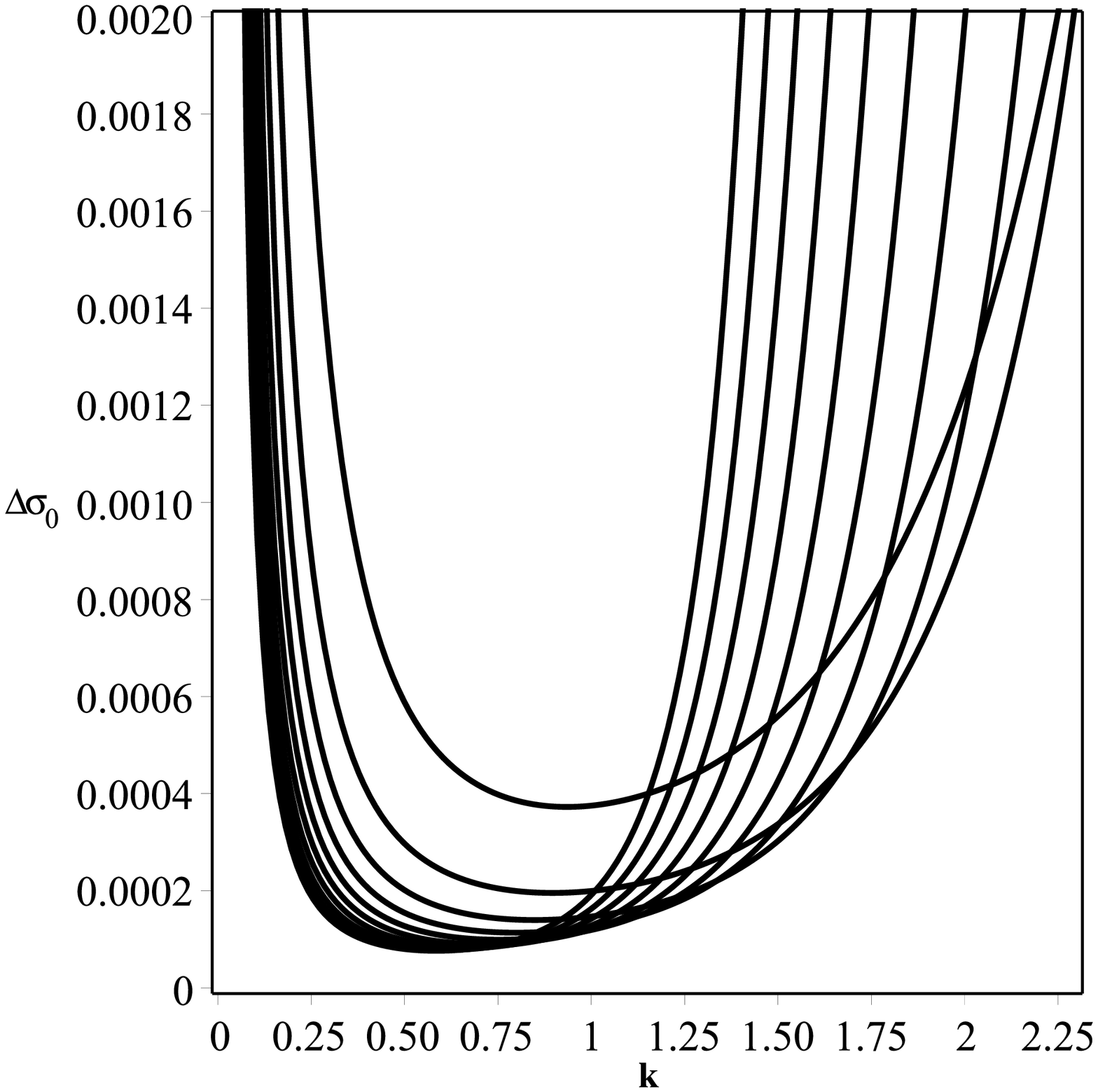} \includegraphics[scale=0.21]{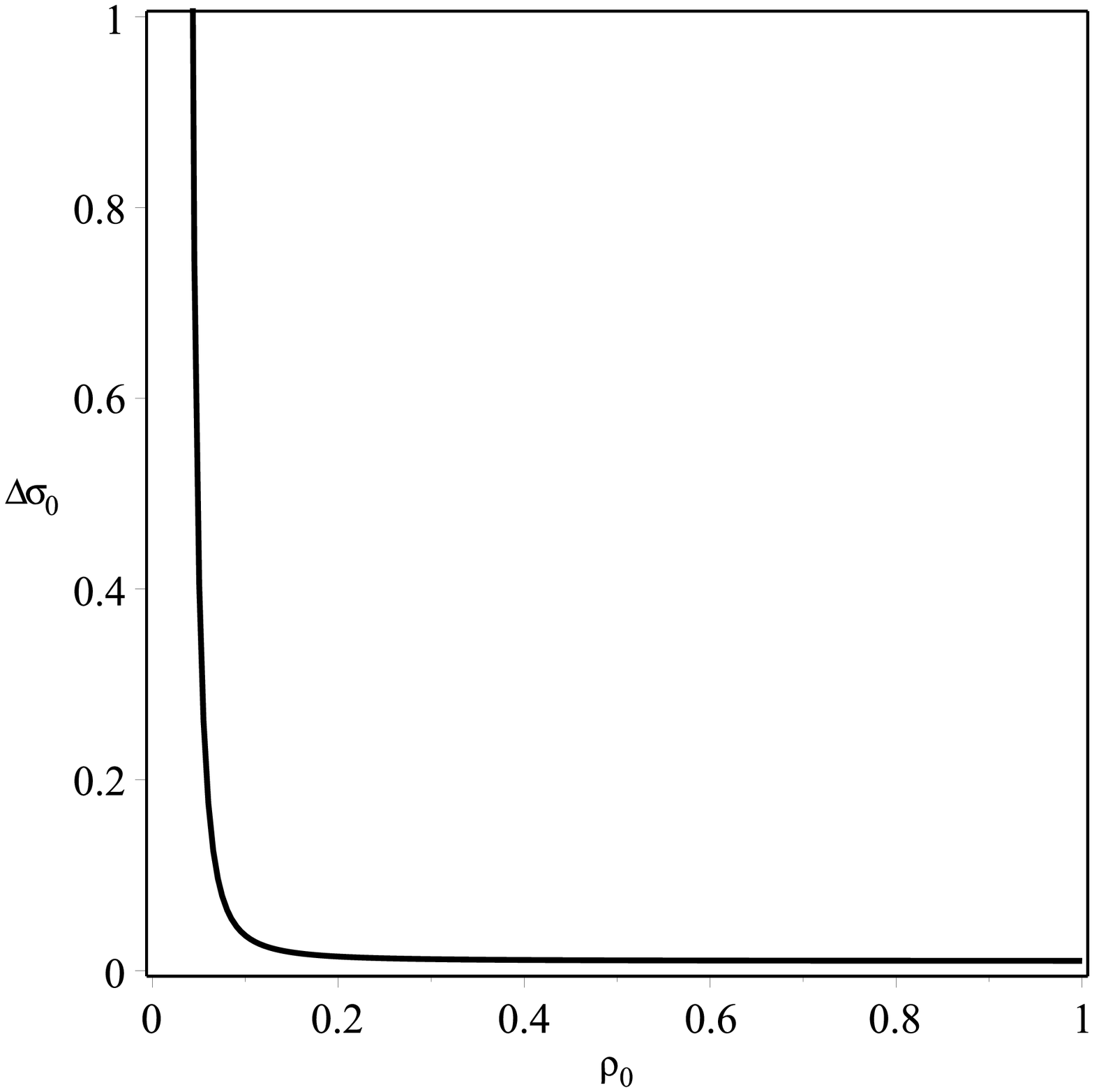}
    \caption{(Left) The optimal bound $\bigtriangleup\sigma_0$ as a function of the momentum $k$ for different values of the Lorentz violation parameter $\sigma_0 = 0.1, ..., 1.0$, where we have fixed $\rho_0 = 1$.  (Right) $\bigtriangleup\sigma_0$ as a function of the parameter $\rho_0$, where we have fixed $k =1$ and $\sigma_0 = 0.1$. Both curves (left) and (right) are plotted for $N = 10^{10}$. }
    \label{fig3}
\end{figure}

 The performance of a given estimator is quantified by the mean square fluctuation, namely $\bigtriangleup\sigma_0$. Thus, the optimal estimator is the one that minimizes $\bigtriangleup\sigma_0$. According to the Cramer-Rao inequality, the optimizing over all the possible quantum measurement provides a lower bound: $(\bigtriangleup\sigma_0)^2 \geq \frac{1}{NH_c(\sigma_0)}$, where $N$ is the number of the identical measurement repeated. By using the Eq. (\ref{F2}), the minimum variance corresponding to the optimal detection of $\sigma_0$ becomes
   \begin{align} \label{Var}
  \bigtriangleup\sigma_0 \geq \sqrt{\frac{(1 - \gamma_k)^2}{N\gamma_k^2}}\frac{\omega^{\mathrm{in}}_k\rho_0}{\pi k^4\sigma_0}\left[\frac{\tanh(\frac{\pi\omega_{-}}{\rho_0})\tanh(\frac{\pi\omega_{+}}{\rho_0})}{\tanh(\frac{\pi\omega_{-}}{\rho_0}) + \tanh(\frac{\pi\omega_{+}}{\rho_0})}\right],
 \end{align}
 which shows the ultimate bound to the precision of the parameter $\sigma_0$ in a particular measurement scheme. The numerical analysis of this expression is summarized in Fig. (\ref{fig3}).

In Fig. (\ref{fig3})(left) we can observe that the minimum bound is obtained for some optimal momenta, which are the values of
$k_{\mathrm{max}}$. We also see from Fig. (\ref{fig3})(left) that the minimum bound decreases as $\sigma_0$ grows. The
numerical data of Table I indicates that the optimal minimum bound approaches $0.76\times 10^{-4}$ as $\sigma_0 \longmapsto 1^{-}$ (Planck scale). This means that near Planck's scale the inequality in Eq. (\ref{Var}) is saturated and the variance $\bigtriangleup\sigma_0$ is as small as possible. Fig. (\ref{fig3})(right) shows the variance $\bigtriangleup\sigma_0$ as a function of the parameter $\rho_0$. Notice that when $\rho_0$ increases, the variance $\bigtriangleup\sigma_0$ approaches its asymptotic minimum ($\bigtriangleup\sigma_0 \rightarrow 0$), indicating a higher precision for estimation. On the other hand, the variance $\bigtriangleup\sigma_0$ diverges when $\rho_0 \rightarrow 0$.

   \begin{table}[ph]
  \caption{Optimal bound $\bigtriangleup\sigma_0$ for different values of $k_{\mathrm{max}}$.}
 {\begin{tabular}{c | c | c | c | c | c}
     \hline
     $k_{\mathrm{max}}$ &  $\sigma_0$ & $\bigtriangleup\sigma_0$ & $k_{\mathrm{max}}$ &  $\sigma_0$ & $\bigtriangleup\sigma_0$ \\
     \hline
     0.934 & 0.1 & $\geq$ $3.73\times 10^{-4}$ & 0.711 & 0.6 & $\geq$ $0.90\times 10^{-4}$  \\
     0.896 & 0.2 & $\geq$ $1.95\times 10^{-4}$ & 0.674 & 0.7 & $\geq$ $0.84\times 10^{-4}$ \\
     0.848 & 0.3 & $\geq$ $1.39\times 10^{-4}$ & 0.641 & 0.8 & $\geq$ $0.81\times 10^{-4}$  \\
     0.799 & 0.4 & $\geq$ $1.14\times 10^{-4}$ & 0.611 & 0.9 & $\geq$ $0.78\times 10^{-4}$ \\
     0.753 & 0.5 & $\geq$ $0.99\times 10^{-4}$ & 0.584 & 1.0 & $\geq$ $0.76\times 10^{-4}$  \\
     \hline
     \end{tabular}}
  \end{table}

In summary, one may observe that the minimum possible bound for the estimation of the Lorentz violation parameter is obtained for some specific value of momentum, $k_{\mathrm{max}}$. Such result can be understood from the fact that $H_c(\sigma_0)$  is maximum for the states with these specific values of momenta. It is also shown that the variance $\bigtriangleup\sigma_0$ approaches its asymptotic minimum as $\sigma_0$ and $\rho$ grows, which indicates a higher precision for estimation of the parameter $\sigma_0$. Now days, our best data sets regarding the early universe come from observation of the cosmic microwave background (CMB)\cite{Mather, Bennett}. Recent experiments have measured the polarization of the CMB at different positions in the sky. An unexpected twist in that polarization would indicate a breakdown of Lorentz invariance in early universe \cite{Kostelecky2, Mewes}. Through the power spectrum $P(k)$ and its derivatives, we can use the Fisher matrix formalism to estimate the accuracy of parameter $\sigma_0$. For instance, the Fisher information matrix for the parameter $\sigma_0$ can be computed as\cite{Tegmark, Takahashi}
\begin{align}
 F_{\sigma_0\sigma_0} = \sum_{k_1, k_2 \sim k_{\mathrm{max}}}\mathrm{Cov}^{-1}(k_1, k_2)\frac{dP(k_1;\sigma_0)}{d\sigma_0}\frac{dP(k_2;\sigma_0)}{d\sigma_0},
\end{align}
where the covariance matrix $\mathrm{Cov}(k_1, k_2)$ between the spectra of $k_1$ and $k_2$ is estimated as $\mathrm{Cov}(k_1, k_2) \equiv \langle (\hat{P}(k_1) - P(k_1))(\hat{P}(k_2) - P(k_2))\rangle$. Here, $P(k) = \langle\hat{P}(k)\rangle$ is the mean power spectrum. Since we focus on the optimal momentum $k_{\mathrm{max}}$, we can treat $\sigma_0$ as a free parameter, and hence the precision of determining $\sigma_0$ for the given power spectrum measurement is given by $\Delta\sigma_0 = (F_{\sigma_0\sigma_0})^{-1/2}$. It is noteworthy to mention that for $k_1 = k_2$ the Fisher information matrix $F_{\sigma_0\sigma_0}$ is equivalent to the calculation of $H_c(\sigma_0)$. Thus, it is possible to improve the precision of the estimation of Lorentz violation parameter by accessing the Fisher information associated to the measurement of the power spectrum or occupation number \cite{Rotondo}.

 \section{Concluding remarks}

In this work, we investigate the estimation of the Lorentz violation parameter by using tools of QET. We have considered a simple toy model of a time dependent deformed dispersion relation to investigate the effects of Lorentz symmetry violation in a  flat spacetime. We calculated the classical part of the QFI as a function of the Lorentz violation parameter $\sigma_{0}$ and the momentum $k$ of the created particles and found that for each $\sigma_{0}$ there is a maximum in a specific value of the momentum, $k_{\mathrm{max}}$. The values of $k_{\mathrm{max}}$ and the respective $\sigma_{0}$ and $\Delta \sigma_{0}$ are disposed in Table I. We found that the ultimate lower bound on the precision of estimation of the parameter $\sigma_0$ is archived for a privileged value of momentum. Our results show that the minimum bound decrease as $\sigma_0$ grows. In addition, they also show that the optimal bound of $\triangle\sigma_0$ saturates when the value of the
parameter $\sigma_0$ approaches to 1 from the left. In this way, we can improve the estimation of the Lorentz violation parameter by choosing particle states with specific values of momentum. Our results imply that the intimate link between Lorentz invariance violation and Planck-scale in early universe affect significantly the performance of the estimation of parameter $\sigma_0$.

We would like to emphasize that although the model for Lorentz
violation we have used to derive our results is far from a full
quantum gravity theory, the results of this paper shed light on the mechanism of
Lorentz symmetry violation in early universe. A complete and
realistic extension of the ideas explored here would be interesting,
but goes beyond the scope of this work. In particular, it would be
interesting to investigate anisotropic effects since the influence
of anisotropy is observed in the spectrum of the CMB \cite{Quercellini, Weinberg}. Another interesting
avenue for future research would be to study the same problem for
Dirac fields, since it is well known that bosonic and fermionic
particle creation processes are qualitatively different. In an
upcoming article we hope to report these issues.






{\bf Acknowledgments}

 HASC would like to thank CAPES (brazilian funding agency) for financial support.

\end{document}